\documentclass[12pt]{article}
\usepackage{graphicx,amssymb,amsfonts,amsmath,epsfig,color}

\newcounter{mnotecount}

\newcommand{\mnotex}[1]
{\protect{\stepcounter{mnotecount}}$^{\mbox{\footnotesize $\bullet$\themnotecount}}$
\marginpar{
\raggedright\tiny\em
$\!\!\!\!\!\!\,\bullet$\themnotecount: #1} }

\makeatletter
\DeclareSymbolFont{AMSb}{U}{msb}{m}{n}
\DeclareSymbolFontAlphabet{\mathbb}{AMSb}
\renewcommand{\section}{\@startsection{section}{1}{\z@}%
                                    {-7ex \@plus -1ex \@minus -.2ex}%
                                    {2.5ex \@plus.2ex}%
                                    {\normalfont\large\scshape\centering}}
\renewcommand{\subsection}{\@startsection{subsection}{2}{\z@}%
                                       {-5ex \@plus -1ex \@minus -.2ex}%
                                       {1.5ex \@plus.2ex}%
                                       {\normalfont\normalsize\scshape}}
\renewcommand{\subsubsection}{\@startsection{subsubsection}{3}{\z@}%
                                       {-5ex \@plus -1ex \@minus -.2ex}%
                                       {1.5ex \@plus.2ex}%
                                       {\normalfont\normalsize\scshape}}

\renewcommand\@seccntformat[1]{\ignorespaces\csname #1name\endcsname\space
                               \csname the#1\endcsname.\quad}   
\newdimen\captionmargin
\newdimen\captionindent
\newdimen\captionwidth
\newcommand{\captionfont}{\slshape}
\newcommand\@captionlabel[1]{\textsc{#1:}\space}
\long\def\@makecaption#1#2{%
  \vskip\abovecaptionskip
  \captionwidth\hsize
  \advance\captionwidth -2\captionmargin
  \sbox\@tempboxa{\@captionlabel{#1}\captionfont #2}%
  \ifdim \wd\@tempboxa >\captionwidth
    \ifdim\captionindent>\z@
      \advance\captionwidth -\captionindent
      \hskip\captionindent
    \fi
    \hskip\captionmargin
    \parbox[t]{\captionwidth}{\leavevmode\hskip-\captionindent
      \@captionlabel{#1}\captionfont #2}%
  \else
    \global \@minipagefalse
    \hb@xt@\hsize{\hfil\box\@tempboxa\hfil}%
  \fi
  \vskip\belowcaptionskip}
\def\eqnarray{%
   \stepcounter{equation}%
   \def\@currentlabel{\p@equation\theequation}%
   \global\@eqnswtrue
   \m@th
   \global\@eqcnt\z@
   \tabskip\@centering
   \let\\\@eqncr
   $$\everycr{}\halign to\displaywidth\bgroup
       \hskip\@centering$\displaystyle\tabskip\z@skip{##}$\@eqnsel
      &\global\@eqcnt\@ne$\;\hfil{##}$\hfil
      &\global\@eqcnt\tw@$\;\displaystyle{##}$\hfil\tabskip\@centering
      &\global\@eqcnt\thr@@ \hb@xt@\z@\bgroup\hss##\egroup
         \tabskip\z@skip
      \cr}
\ifcase \@ptsize
\or
\or
\fi
  \divide\@tempdima\baselineskip
  \@tempcnta=\@tempdima
\makeatother
\begin{document}

\renewcommand{\theequation}{\arabic{section}.\arabic{equation}}
\renewcommand{\thefigure}{\arabic{figure}}
\newcommand{\gapprox}{%
\mathrel{%
\setbox0=\hbox{$>$}\raise0.6ex\copy0\kern-\wd0\lower0.65ex\hbox{$\sim$}}}
\textwidth 165mm \textheight 220mm \topmargin 0pt \oddsidemargin 2mm
\def\ib{{\bar \imath}}
\def\jb{{\bar \jmath}}

\newcommand{\ft}[2]{{\textstyle\frac{#1}{#2}}}
\newcommand{\be}{\begin{equation}}
\newcommand{\ee}{\end{equation}}
\newcommand{\bea}{\begin{eqnarray}}
\newcommand{\eea}{\end{eqnarray}}
\newcommand{\Identity}{{1\!\rm l}}
\newcommand{\cx}{\overset{\circ}{x}_2}
\def\CN{$\mathcal{N}$}
\def\CH{$\mathcal{H}$}
\def\hg{\hat{g}}
\newcommand{\bref}[1]{(\ref{#1})}
\def\espai{\;\;\;\;\;\;}
\def\zespai{\;\;\;\;}
\def\avall{\vspace{0.5cm}}
\newtheorem{theorem}{Theorem}
\newtheorem{acknowledgement}{Acknowledgment}
\newtheorem{algorithm}{Algorithm}
\newtheorem{axiom}{Axiom}
\newtheorem{case}{Case}
\newtheorem{claim}{Claim}
\newtheorem{conclusion}{Conclusion}
\newtheorem{condition}{Condition}
\newtheorem{conjecture}{Conjecture}
\newtheorem{corollary}{Corollary}
\newtheorem{criterion}{Criterion}
\newtheorem{defi}{Definition}
\newtheorem{example}{Example}
\newtheorem{exercise}{Exercise}
\newtheorem{lemma}{Lemma}
\newtheorem{notation}{Notation}
\newtheorem{problem}{Problem}
\newtheorem{prop}{Proposition}
\newtheorem{rem}{{\it Remark}}
\newtheorem{solution}{Solution}
\newtheorem{summary}{Summary}
\numberwithin{equation}{section}
\newenvironment{pf}[1][Proof]{\noindent{\it {#1.}} }{\ \rule{0.5em}{0.5em}}
\newenvironment{ex}[1][Example]{\noindent{\it {#1.}}}

\thispagestyle{empty}


\begin{center}

{\LARGE\scshape Particle production from marginally trapped surfaces of general
spacetimes \par}
\vskip15mm

\textsc{Jos\'{e} M. M. Senovilla\footnote{E-mail: josemm.senovilla@ehu.es}
}
\par\bigskip
{\em
F\'{i}sica Te\'{o}rica, Universidad del Pa\'{\i}s Vasco, Apartado 644, 48080 Bilbao, Spain.}\\[.1cm]
\par\bigskip

\textsc{Ram\'{o}n Torres\footnote{E-mail: ramon.torres-herrera@upc.edu}
}
\par\bigskip
{\em
Dept. de F\'{i}sica Aplicada, Universitat Polit\`{e}cnica de Catalunya, Barcelona, Spain.}\\[.1cm]

\vspace{5mm}

\end{center}

\begin{abstract}
We provide a general formalism that allows to analyze the phenomenon of tunneling in arbitrary spacetimes. We show that a flux of particles produced by tunneling through general marginally trapped surfaces may be perceived by some privileged observers. We discuss how this particle perception can be related to Hawking/Unruh radiation in specific cases. Our approach naturally leads to an expression for the effective surface gravity of marginally trapped surfaces. The procedure is applicable to general astrophysical and cosmological dynamical situations. Some practical examples for known and new cases are provided.
\end{abstract}




\setcounter{equation}{0}

\section{Introduction}
In 1975 Hawking showed \cite{Haw75} that black holes (BH) radiate a thermal spectrum of particles by deriving an exact expression
for their entropy and temperature. This celebrated result was based on quantum field theory on a fixed curved stationary background (Schwarzschild's solution). In the same work Hawking used a heuristic picture in order to explain in physical terms the existence of radiation from stationary black holes. In this picture a pair of virtual particles is created and one of the members of the pair manages to tunnel through the event
horizon.
Then, the member of the pair with positive $E$ ---where $E$ is the constant of motion of the particles associated to the static Killing vector, so that it corresponds to the energy as measured by static observers at infinity--- is emitted from the black hole while the member with negative $E$ falls into the black hole.
This possibility is based on the fact that for stationary black holes the event horizon is a {\em Killing horizon} so that a Killing vector becomes null at the event horizon, changing its causal character from timelike to spacelike across it. This
allows for the existence of negative $E$ states in the black hole interior.

On realistic grounds, one expects stationary black holes to be very rare in the Universe. Actual black holes are subject to dynamical processes such as their formation, the accretion of matter/energy and their own back-reaction to the emission of Hawking radiation. Moreover, due to quantum gravity effects the event horizon could either not exist or be a meaningless concept \cite{Haji}\cite{Ash}, see \cite{BCGJ} ---and references therein--- for a recent discussion.
The questions that arise are then whether there is \textit{Hawking radiation} when there could not even be an event horizon and, if so, which is the correct tunneling horizon associated with the phenomenon.

Herein, we investigate if the answer to these questions could come from the heuristic tunneling picture suggested by Hawking. Indeed, some semiclassical methods have been proposed in recent years in order to effectively compute \textit{Hawking radiation} in agreement with the picture, most remarkably the so called \textit{Hamilton-Jacobi method} \cite{Sri&Padma} and the \textit{null geodesic method} \cite{P&W}. The mere existence of these methods indicates that the tunneling could be more than just a \emph{heuristic} picture. In fact, it has been recently shown that the tunneling approach can be justified from the point of view of proper Quantum Field Theory in curved spacetime
if one assumes the local existence of a Killing horizon \cite{Moretti}.

While originally developed in order to shed some light on Hawking radiation from stationary black holes, the tunneling methods have been recently modified in such a way that {\em some} dynamical situations can be treated (see, for example, the review \cite{revVz}). In particular,  by working with dynamical spherically symmetric spacetimes and using their Kodama vector field \cite{kodama} to replace the nonexistent Killing vector, it has been possible to associate particle production with the existence of the apparent 3-horizon, i.e. the spherically symmetric marginally trapped tube (which is the unique spherically symmetric hypersurface foliated by marginally trapped surfaces \cite{B&S}, actually {\em round spheres}) in accordance with the old proposal in \cite{Haji}, as well as to provide some properties of that radiation.

Our aim in this paper is to analyze the perception of particle production by tunneling in general spacetimes. We argue that, according to some observers, the radiation may exist associated with generic marginally trapped surfaces whenever they separate, locally, trapped from untrapped surfaces. This ``boundary'' property is related to the important concept of (local) outermost stability \cite{AMS,AMS1} of marginally trapped surfaces. As an important remark, notice that there can be many different marginally trapped surfaces in a spacetime and the marginally trapped tubes that they form interweave each other in very complicated ways \cite{AG,B&S}. We will argue that the perception of particle production by tunneling is associated with {\em all of them}. As an extreme example in spherically symmetric spacetimes, non-spherically symmetric marginally trapped surfaces with points beyond the apparent 3-horizon, and even reaching flat portions of the space-time, exist \cite{B&S0} and we argue that, as perceived by some specific observers, they have associated particle production.
Whether or not the radiation ends up reaching future asymptotic regions is quite another matter, though, that depends on the non-local structure of the space-time. Many other interesting implications follow from the idea that particle production is based on tunneling through (separating) marginally trapped surfaces.

Throughout this paper, and when referring to our own results, we have restricted the expression \textit{Hawking radiation} to the tunneling radiation observed by stationary observers in stationary asymptotically flat spacetimes.
Whenever this meaning of \textit{Hawking radiation} does not apply, we have instead used \textit{particle production} if there are tunneled particles perceived by specific observers.

The plan of the paper is
as follows. Section \ref{GP} is devoted to the geometrical preliminaries used in order to describe general surfaces and some of their properties. Section \ref{Tunneling} first describes the usual Hamilton-Jacobi method used to analyze Hawking radiation as due to the tunneling of null particles through a horizon. Then, the method is generalized to arbitrary spacetimes containing marginally trapped surfaces in subsection \ref{GITF}. Some practical examples are worked out in section \ref{SE}. Finally, our results are discussed in section \ref{Conclu}.

\section{Geometrical Preliminaries}\label{GP}

Let $(\mathcal V, g)$ be a 4-dimensional causally orientable spacetime with metric signature $\{-,+,+,+\}$ and with local coordinates $\{x^\alpha\}$.
Let $\mathcal S$ denote a connected 2-dimensional surface with local intrinsic coordinates $\{\lambda^A\}$
imbedded in $\mathcal V$ by the $C^3$ parametric equations
\[
x^\alpha=\Phi^\alpha(\lambda^A).
\]
The tangent vectors $\vec{e}_A$ of $\mathcal S$  are locally given by
\[
\vec{e}_A\equiv e^\mu_A \frac{\partial}{\partial x^\mu}\rfloor_{\mathcal S} \equiv \frac{\partial \Phi^\mu}{\partial \lambda^A} \frac{\partial}{\partial x^\mu}\rfloor_{\mathcal S}
\]
so that the first fundamental form of $\mathcal S$ in $\mathcal V$ is
\[
\gamma_{AB}\equiv g_{\mu\nu}\rfloor_{\mathcal S} \frac{\partial \Phi^\mu}{\partial \lambda^A}\frac{\partial \Phi^\nu}{\partial \lambda^B}.
\]
We are interested in spacelike surfaces $\mathcal S$ in which case $\gamma_{AB}$ is positive definite. The two linearly independent null future-directed one-forms normal to $\mathcal S$ are denoted by $l^\pm_\mu$. They satisfy
\[
l^\pm_\mu e^\mu_A=0,\ \ \ l^+_\mu l^{+\mu}=0, \ \ \ l^-_\mu l^{-\mu}=0
\]
and, without loss of generality, we choose them to satisfy the convenient normalization condition
\begin{equation}\label{norm}
l^+_\mu l^{-\mu}=-1.
\end{equation}
The covariant derivatives on $(\mathcal V, g)$ and on $(\mathcal{S}, \gamma)$ are related through \cite{Kriele}\cite{ONeill}
\[
e^{\rho}_{A}\nabla_{\rho}e^{\mu}_{B}=\overline{\Gamma}^C_{AB} e^{\mu}_{C}-K^{\mu}_{AB}
\]
where $\overline{\Gamma}^C_{AB}$ are the coefficients of the Levi-Civita connection $\overline{\nabla}$ of $\gamma$ (so that $\overline{\nabla}\gamma=0$) and $K^{\mu}_{AB}$ is the shape tensor (also called second fundamental form vector or extrinsic curvature vector) of $\mathcal S$  in $(\mathcal V, g)$. The shape tensor is normal to ${\cal S}$ and thus it can decomposed as
\[
K^{\mu}_{AB}=-K^-_{AB} l^{+\mu} -K^+_{AB} l^{-\mu},
\]
where $K^\pm_{AB}$ are called the two null (future) second fundamental forms of $\mathcal S$  in  $(\mathcal V, g)$
given by
\[
K^\pm_{AB}\equiv e^\nu_A e^\mu_B \nabla_\nu l^\pm_\mu , \hspace{2cm} K^\pm_{AB}=K^\pm_{BA} \, .
\]
The mean curvature vector of $\mathcal S$  in $(\mathcal V, g)$ \cite{Kriele}\cite{ONeill}
is the trace of the shape tensor
\[
H^{\mu}\equiv \gamma^{AB} K^{\mu}_{AB}
\]
where $\gamma^{AB}$ is the contravariant metric on $\mathcal S$  ($\gamma^{AC} \gamma_{CB}=\delta^A_B$).
Clearly, the mean curvature vector is orthogonal to $\mathcal S$ and can be written in the form
\[
H^{\mu}= -\theta^- l^{+\mu} -\theta^+ l^{-\mu},
\]
where
\[
\theta^\pm\equiv \gamma^{AB} K^{\pm}_{AB}
\]
are the traces of the null second fundamental forms, also called the (future) null expansions.
We will be specially interested in surfaces in which $\vec{H}\ (:\neq \vec{0})$ is null everywhere on $\mathcal S$, keeping its causal orientation (future or past) and pointing consistently along one of two null directions $l^\pm$. These surfaces are called marginally (future or past) trapped surfaces (MTS) and satisfy either $\{\theta^+=0, \theta^- \leq 0\}$ or $\{\theta^-=0, \theta^+ \leq 0\}$ for the future case (reverse inequalities for the past case). For concreteness, and unless stated otherwise, from now on we will tacitly assume that we are dealing with the first case when considering a MTS. We will also assume that the non-vanishing expansion is strictly negative for simplicity and to avoid unnecessary complications.

The {\em unique} vector field dual to $\vec{H}$ in the plane orthogonal to $\mathcal S$, called the \textit{dual expansion vector} \cite{Tung}\cite{SenoProc}, takes the form
\begin{equation}
\ast H^{\mu}= -\theta^- l^{+\mu} +\theta^+ l^{-\mu}.
\end{equation}
This vector field defines the (generically unique) direction with vanishing expansion of $\mathcal S$ \cite{Tung}\cite{SenoProc}.
$\ast \vec H$ is timelike for untrapped surfaces, spacelike for trapped surfaces and null (equal to $\vec H$) for MTS.
Moreover, in the framework of quasi-local Hamiltonians it defines the direction of a Hamiltonian flow at $\mathcal S$ \cite{Tung}\cite{Anco} (when $\mathcal S$ is compact and under some mild assumptions, this flow leads to the expression for the \textit{Hawking energy} enclosed by $\mathcal S$). In particular, in spherically symmetric spacetimes and when $\mathcal S$ is chosen to be a round sphere, $\ast \vec{H}$ is parallel to the \textit{Kodama vector} \cite{kodama}. For our purposes, it is enough to note that, for untrapped surfaces,
\begin{equation}\label{GKO}
\hat\zeta\ \equiv \frac{\ast\vec{H}}{\sqrt{- g(\ast\vec{H},\ast\vec{H})}}
\end{equation}
defines on $\mathcal S$ the 4-velocity of privileged observers with respect to $\mathcal S$, in the sense that they measure no expansion of ${\cal S}$.

\section{Tunneling}\label{Tunneling}
Let us now consider the perception of particle production
according to the tunneling approach. Specifically, we will use the tunneling procedure called the \textit{Hamilton-Jacobi method} \cite{Sri&Padma}\cite{revVz} that is based on the usual \textit{complex path method} from the theory of semiclassical approximations.
Briefly, the standard use of this method considers the Klein-Gordon equation for the wave function $\Phi$ describing the massless particle that tunnels through a \textit{horizon} (typically a BH horizon). Then, the usual formal substitution $\Phi=\exp (- i S/\hbar)$ is performed. By expanding $S$  in powers of $\hbar/i$ ($S=S_0+ (\hbar/i) S_1+\ldots$) and neglecting the terms of order ($\hbar/i$) and greater that arise in the Klein-Gordon equation, one obtains
\begin{equation}\label{HJ}
S_{0,\alpha} S_{0,\beta} g^{\alpha\beta}=0
\end{equation}
that is to say, the Hamilton-Jacobi equation for the action $S_0$ corresponding to the massless particle.
On the other hand, the action can be obtained by means of
\begin{equation}\label{S0}
S_0 (x^\alpha)=\int_{x_{0}}^{x} dS_{0} = \int_{x^\alpha_0}^{x^\alpha} S_{0,\mu} dx^\mu,
\end{equation}
where the integration follows a null geodesic path and one must use a well-behaved coordinate system describing the whole path from its initial point $x_{0}\in {\cal V}$ to its final point $x\in {\cal V}$.

In this approximation, classically forbidden trajectories can be treated by letting the action acquire an imaginary part \cite{Landau}\cite{Sri&Padma}. For example, Quantum Mechanics tells us that the wave function just outside a barrier can be straightforwardly written as $\Phi_{out}\sim\Phi_{in} \exp\{- \mbox{Im} S_0/\hbar\}$ (where $\Phi_{in}$ is the wave function at the barrier's entrance and Im$S_0$ is the imaginary part of the action with integration limits $x^\alpha_{in}$ and $x^\alpha_{out}$), so that the escape probability $\Gamma$ of the particle will approximately be
\[
\Gamma=\frac{|\Phi_{out}|^2}{|\Phi_{in}|^2} \sim \exp\{- (2/\hbar) \mbox{Im} S_0\}.
\]
By taking into account (\ref{S0}) one checks that this probability is coordinate independent, as expected.
Among the classically forbidden situations in which $S_0$ acquires an imaginary part we can include the case of a particle tunneling through the event horizon of a black hole, what was first shown for the Schwarzschild black hole (see \cite{Sri&Padma}\cite{P&W}).

If we want to consider the Hamilton-Jacobi approach to describe particle production a clear justification of the used approximations is necessary.
It is known that these approximations can only be justified if the tunneling massless particle has a high frequency. 
For the original Schwarzschild black hole and the tunneling of particles through its event horizon this is, indeed, the case: the external stationary observers associated with the Killing vector field $\vec \xi$ that becomes null at the horizon have a normalized 4-vector that we denote by
$\hat\xi\ (\equiv\vec{\xi}/\sqrt{-\xi_{\mu}\xi^{\mu}}\ )$. These observers measure an energy $\hat E$ of a particle reaching them given by
\[
\hat E=-p_\alpha \hat\xi^\alpha = \frac{E}{\sqrt{-\xi_{\mu}\xi^{\mu}}},
\]
where $E=-p_\alpha \xi^\alpha$ is the energy of the particle as measured by the stationary observers at infinity. Note that the expression above, and therefore the measured energy  $\hat E$, diverges as the stationary observer approaches the event horizon where $\vec{\xi}$ becomes a light-like vector. In other words, when a static observer at infinity observes a massless particle with energy $E$, no matter how small this could be, she must deduce that its energy
after tunneling was very high and that, during its travel, its frequency has been redshifted by the effect of the gravitational field.

As commented in the introduction, some steps have already been taken in order to generalize the tunneling methods beyond its original application to Schwarzschild black holes. In particular, the methods have been extended to spherically symmetric dynamical black hole solutions. The basic ingredient for this type of dynamical situations (see, for example, the review \cite{revVz}) was the replacement of the now absent static Killing vector
by the Kodama vector $K^{\mu}$. This is utilized to define a \textit{Kodama ``energy''} $\omega \equiv - S_{0,\alpha} K^\alpha$ that is assumed to be a finite and non-vanishing function
thereby showing that the action of the particle acquires an imaginary part when traversing the hypersurface $r=2m$, where $m$ is the mass function \cite{M&S}\cite{Hayward} and $r$ is called the \textit{areal radius} since the area of the
round spheres is $4\pi r^2$. The hypersurface $r=2m$ is not, generally speaking, the event horizon, but the apparent 3-horizon (A3H),
see e.g. \cite{B&S}.

\subsection{General invariant tunneling formalism}\label{GITF}
In what follows, we show that tunneling radiation can be associated with marginally trapped surfaces in general spacetimes.
As we argued in section \ref{GP}, for a general surface $\mathcal S$ in a general spacetime the generalization of the Kodama vector
is (except for a non-null multiplicative factor)
the dual expansion vector $\ast\vec{H}$. Consequently, by analogy we are going to assume that, in the general case, the invariant
\begin{equation}
\tilde\omega \equiv - S_{0,\alpha} \ast\! \! H^\alpha \label{energy}
\end{equation}
is finite and vanishing nowhere \footnote{In other words, we are demanding that the privileged observers associated with the dual expansion vector of untrapped surfaces cannot declare that a measured massless particle has no energy.}.

Given any MTS ${\cal S}$, consider a local foliation $\mathcal N (t)$ of the spacetime by pieces of spacelike hypersurfaces in such a way that, for a given value $t_{0}$ of $t$, our MTS ${\cal S} \subset {\cal N}(t_{0})$. Construct then a local tube foliated by spacelike surfaces $\mathcal S(t)$, each of them lying in one of the $\mathcal N (t)$ ($\mathcal S(t_0)\ \equiv \mathcal S$).
\begin{figure}
\centering
\includegraphics[scale=.9]{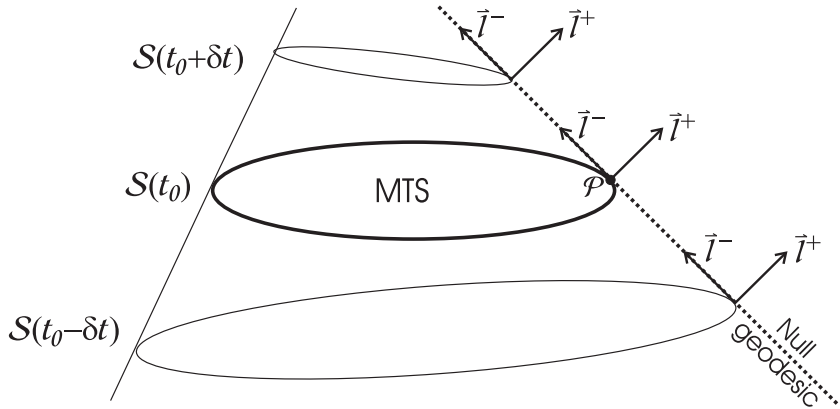}
\caption{\label{NS} A schematic representation showing some surfaces from the tube. The lightlike geodesic with tangent vector $\vec l^-$ is described by a dashed line. For the purposes of  this figure we are considering the case of compact surfaces $\mathcal S (t)$.}
\end{figure}
In order to analyze the tunneling of massless particles across a point $\cal P\in {\cal S}$, the tube is further constrained so that it contains the null geodesic generated at ${\cal P}$ by the tangent vector $\vec{l}^-$ normal to $\mathcal S(t_0)$ (see figure \ref{NS}), and such that each ${\cal S}(t)$ has a negative (respectively positive) expansion $\theta^{+}$ at their intersection with the null geodesic for $t>t_{0}$ (resp.\ for $t< t_{0}$). As mentioned before, this is related to the outermost stability (or instability) of ${\cal S}$ \cite{AMS,AMS1}.

Define a local basis $\{\vec{l}^+,\vec{l}^-,\vec{e}_1,\vec{e}_2\}$ along the null geodesic by letting $\vec l^{\pm}$ be the future null normals of the $\{{\cal S}(t)\}$ family and $\vec{e}_{1,2}$  spacelike orthonormal tangent vector fields on each ${\cal S}(t)$.
The metric tensor can then be written as
\[
g_{\alpha\beta}=-l^+_\alpha l^-_\beta -l^-_\alpha l^+_\beta+e_{1\alpha} e_{1\beta}+e_{2\alpha} e_{2\beta}
\]
and the expressions (\ref{HJ}) and (\ref{energy}) take, respectively, the form
\begin{eqnarray}
0 &=&-2 (l^{+\alpha} S_{0,\alpha}) (l^{-\beta} S_{0,\beta})+(e_1^\alpha S_{0,\alpha})^2+(e_2^\alpha S_{0,\alpha})^2, \label{system0}\\
\tilde\omega &=& \theta^- (l^{+\alpha} S_{0,\alpha}) - \theta^+ (l^{-\alpha} S_{0,\alpha}).
\label{system}
\end{eqnarray}
By considering the above expressions as two simultaneous equations for $l^{\pm\alpha} S_{0,\alpha}$ one gets for particles trying to cross the MTS
\[
l^{-\alpha} S_{0,\alpha}=-\frac{\tilde\omega+\sqrt{\tilde\omega^2+2 \theta^+ \theta^- [(e_1^\alpha S_{0,\alpha})^2+(e_2^\alpha S_{0,\alpha})^2]}}{2 \theta^+}.
\]
Taking into account that
$\theta^+\rfloor_{MTS} = 0$ on the marginally trapped surfaces under consideration,
this inform us that $l^{-\alpha} S_{0,\alpha}$ diverges as the particle crosses the MTS (in other words, the MTS acts as a classical impenetrable barrier for the particle). As a consequence, when the particle approaches the marginally trapped surface $S_{0,\mu}$ can be written in our local basis as $- (l^{-\alpha} S_{0,\alpha}) l^+_\mu$ plus `non-divergent terms'. Hence, when  a lightlike particle tunnels from a point on the \textit{interior} (`\textit{In}') of the MTS to a point on its \textit{exterior} (`\textit{Out}') one has
\begin{equation}
\mbox{Im} S_0=\mbox{Im} \int_{In}^{Out} S_{0,\mu} dx^\mu=- \mbox{Im} \int_{In}^{Out} (l^{-\alpha} S_{0,\alpha})\ \mathbf{\underline{l}^+},\label{iml+}
\end{equation}
where we have introduced the 1-form $\mathbf{\underline{l}^+}\equiv l^+_\alpha dx^\alpha$ and we have taken into account that the only term that contributes to the imaginary part of $S_{0}$ is the one diverging as the MTS is crossed ---a fact that requires regularizing the integral according to Feynman's $i\epsilon$-prescription.
Expression (\ref{iml+}) can be rewritten in order to make its divergence on the MTS explicit by noticing that, as the null geodesic approaches the MTS,
$l^{-\alpha} S_{0,\alpha} \approx -\tilde\omega/\theta^+$ and, thus,
\begin{equation}\label{ims0}
\mbox{Im} S_0= \mbox{Im} \int_{In}^{Out} \frac{\tilde\omega}{\theta^+} \ \mathbf{\underline{l}^+}.
\end{equation}

With regard to the integration path, tunneling only occurs if the path of the particle across the point on the MTS is classically forbidden, what provide us with an imaginary part for the action. Or course, there are many \emph{classically} allowed paths for light-like particles
crossing any MTS,
but then Im$S_0=0$ along that path and we will be simply dealing with a \textit{travelling particle} rather than with a tunneling particle.
The action $S_{0}$ will only acquire an imaginary part if the particle follows a non-classical path in order to cross the MTS.
The natural directions to cross the MTS following a null path are those defined by the null vectors $\vec l^\pm$ normal to $\cal S$, but only $\vec l^-$ gives a non-zero  result for the integral. Therefore, our prescription to evaluate (\ref{ims0}) is to follow the direction defined by $\vec l^-$ when the integral is evaluated around the point in the MTS. Of course, the validity of this prescription must be checked by the correctness of the results obtained through its use.

Now, denote by $\lambda$ the parameter along the integration path, and recall that $ l^+_{\mu}l^{-\mu}=-1$. Close to the point on the MTS $\theta^+\simeq d\theta^+/d\lambda\rfloor_{MTS} (\lambda-\lambda_{MTS})$ so that Feynman's $i\epsilon$-prescription provides
\begin{equation}\label{pathint}
\mbox{Im} S_0= -\mbox{Im} \int_{In}^{Out} \frac{\tilde\omega}{d\theta^+/d\lambda\rfloor_{MTS} (\lambda-\lambda_{MTS}-i\epsilon)} \ d\lambda =\left.\frac{\pi\tilde\omega}{\tilde\kappa}\right\rfloor_{MTS},
\end{equation}
where
\[
\tilde\kappa\equiv -l^{-\alpha}\nabla_\alpha \theta^+.
\]
From this we can, for example, compute the exponential part of the semiclassical emission rate as
\[
\Gamma \sim \exp (-2\ \mbox{Im} S_0)=\exp \left(-2 \pi\left.\frac{\tilde\omega}{\tilde\kappa}\right\rfloor_{MTS}\right).
\]

Observe that in the previous computations we have been using the approximation in which the  Hamilton-Jacobi equation is applicable in order to describe a massless particle tunneling through a MTS. We will now justify that the particle has a high enough frequency to admit such a description.
In order to see that this is, indeed, the case consider as an example a smooth normalized future-directed timelike vector field $\vec w$ along the trajectory of the tunneling particle  and satisfying, as it approaches the MTS, $\lim \vec w=\hat\zeta$. We will consider this vector field as the 4-velocity of our fiducial observers along the trajectory.
According to (\ref{GKO}), fiducial observers approaching the MTS measure an energy for the particle $\hat E=-S_{0,\alpha} w^\alpha$ satisfying
\[
\lim_{P\rightarrow P_{MTS}} \hat E= \frac{\tilde\omega_{MTS}}{\sqrt{- g(\ast\vec{H},\ast\vec{H})_{MTS}}}=\infty
\]
since, in this case, $\ast\vec{H}$ becomes a light-like vector. In this way, our fiducial observers close enough to the MTS will measure an arbitrarily high energy for the emitted particle as they approach the MTS.

In generic situations, marginally trapped surfaces belong to marginally trapped tubes (actually, they belong to {\em many} such tubes! \cite{AG,AMS1,B&S}).\footnote{Furthermore, they classically persist as such for some time in the space-time \cite{AMS2} ---in other words, for a given foliation $\{{\cal N}(t)\}$ one can find a MTS on each leaf provided there is at least one such MTS and the null energy condition is not violated.}  This implies that the region surrounding any given MTS is plagued with many other MTSs. According to our discussion, each of these MTS produces a certain amount of radiation. Of course, the radiation associated to any particular MTS will only be visible to some of the observers who enter, or come extremely close to, the causal future of the given MTS. In particular, to measure a portion of radiation at far-away asymptotic regions one needs to consider non-local properties of the space-time.

\subsection{Surface gravity and temperature for compact MTS}\label{compactMTS}

In the case of a compact surface ${\cal S}$ we can define the vector
\[
\vec\zeta\equiv\sqrt{\frac{A(\mathcal S)}{16\pi}}(*\vec H),
\]
where $A(\mathcal S)$ stands for the area of the surface $\mathcal S$.
This vector does not only determines the direction of the Hamiltonian flow ---as the dual expansion vector does (sect.\ref{GP})--- but it can be considered as \emph{the} Hamiltonian flow for compact surfaces. In particular, in spherically symmetric spacetimes and when $\mathcal S$ is chosen to be a {\em round} 2-sphere, $\vec\zeta$ coincides with the \textit{Kodama vector} \cite{kodama}. This is why $\vec \zeta$ can be considered as a generalization of the Kodama vector for compact surfaces in general spacetimes.

For a given massless particle one can define the quantity
\begin{equation}\label{EEnergy}
\omega\equiv -S_{0,\alpha} \zeta^\alpha
\end{equation}
which has dimensions of energy. We will call it the \textit{effective energy} of the particle. It is, in fact, a regularized energy since it is straightforwardly related to the energy measured by our preferred observers (\ref{GKO}); however, it does not diverge on the MTS.
On the other hand, the Hamilton-Jacobi equation (\ref{HJ}) inform us that $dS$ is lightlike. In this way, for untrapped surfaces ($\ast \vec H$ timelike) the effective energy can only be positive. However, for trapped surfaces ($\ast \vec H$ spacelike) both positive and negative values of $\omega$ are allowed for the particle. This is a crucial fact for the interpretation of the tunneling since it permits a virtual pair of particles to become real when, after the tunneling occurs, on one side of the MTS the $\omega>0$ particle is allowed to cross a family of untrapped surfaces $\mathcal S(t)$ while, on the other side, the $\omega<0$ particle travels through a family of trapped surfaces.

Since the emission rate for a static black hole takes the form $\Gamma \sim \exp (-2 \pi E/\kappa\rfloor_{MTS})$, where $\kappa$ is the surface gravity of the horizon, this suggests rewriting the general emission rate for the case of compact marginally trapped surfaces as
\[
\Gamma \sim \exp \left(-2 \pi\left.\frac{\omega}{\kappa}\right\rfloor_{MTS}\right),
\]
where
\begin{equation}\label{ESG}
\kappa\equiv -\sqrt{\frac{A(\mathcal S_{MTS})}{16\pi}}\  l^{-\alpha}\nabla_\alpha \theta^+\rfloor_{MTS}
\end{equation}
can be considered as the \textit{effective surface gravity} associated with a compact (future) MTS included in the family of chosen surfaces $\mathcal S (t)$.\footnote{One can trivially rewrite this for the case of \emph{past} MTS. See subsection \ref{BI}, expression (\ref{kappapast}).} Important properties of this quantity are that it is geometrically invariant as well as independent of the parametrization of $\vec l^\pm$. This last property follows from the vanishing of $\theta^+$ on the MTS.\footnote{Of course, formula (\ref{ESG}) is correct for $\vec l^\pm$ subject to the normalization condition (\ref{norm}). If one insists in letting the parametrization of $\vec l^\pm$ free, the minus sign in Eq.(\ref{ESG}) should be replaced by $(l^-_\mu l^{+\mu})^{-1}$.} It is also remarkable that this fact is a direct consequence of our previous selected prescription for the integration path, a fact that reinforces this choice.

Moreover, if our interpretation above of the tunneling phenomenon is correct the tunneling particle with $\omega<0$ should tunnel the (future) MTS following an ingoing integration path with tangent vector $\vec l^-$. During its trajectory a family of surfaces $\mathcal S(t)$ are traversed with expansion that are first positive, then zero and finally negative resulting in $\kappa\geq 0$: a very desirable result.

The existence of an effective surface gravity suggests the existence of a \textit{Temperature} for the MTS. However, care must be taken in the following comments since it is only appropriate to talk about a temperature if the MTS is part of a marginally trapped tube with slowly varying effective surface gravity, see in this respect \cite{B,B1,Barc}. In the general case of \textit{thermal emission} the emission rate can be written as $\Gamma \sim \exp (- E/T)$, where $E$ is the energy of the emitted particle and $T$ is the temperature of the radiation. This clearly suggests defining an \textit{effective temperature} of the compact MTS for the chosen family of surfaces $\{\mathcal S(t)\}$ as
\[
T\equiv\frac{\kappa}{2 \pi}.
\]

Some particular cases of the expression (\ref{ESG}) have been already found by different means and are well-known. For example, when the surfaces are round spheres in the Schwarzschild spacetime it gives the correct expected surface gravity for the horizon ($\kappa=(4 M)^{-1}$, where $M$ is the mass of the black hole), usually obtained by using its timelike Killing vector $\vec\chi$ and $\chi^\beta\nabla_\beta\chi^\alpha=\kappa \chi^\alpha$ on the horizon \cite{Wald}. Similarly, a  definition of surface gravity valid only for spherically symmetric spacetimes was put forward by Hayward in \cite{Hkappa} using the Kodama vector $\vec K$ and $K^\beta\nabla_{[\beta} K_{\alpha]}=-\kappa K_\alpha$. For this case, i.e. when we particularize our $\kappa$ for MTSs that are round spheres in spherically symmetric spacetimes, our expression also coincides with Hayward's expression\footnote{See the first example in section \ref{SE}.}.

\section{Some examples}\label{SE}
In this section we use our formalism to compute particle production (according to our privileged observers) in some spacetimes possessing marginally trapped surfaces. Some of them are already well known but are included simply to check the formalism and for comparison, while other examples are treated here for the first time.

\subsection{Dynamical Spherically Symmetric Black Hole}

The general spherically symmetric metric in advanced Eddington-Filkenstein coordinates takes the form
\begin{equation}\label{SEEF}
ds^2=-e^{2 \beta} \left( 1-\frac{2 m}{r}\right) du^2+2 e^{\beta} du dr+ r^2 (d\vartheta^2+\sin^2\vartheta d\varphi^2),
\end{equation}
where $\beta=\beta(u,r)$ and $m=m(u,r)$.
If we want to describe particle production associated with the family of round spheres we have to consider the null vector fields normal to them, i.e., the radial null vectors that can be written as
\begin{equation}\label{ltd}
\vec l^+=\frac{\partial}{\partial u}+ \frac{e^{\beta}}{2} \left( 1-\frac{2 m}{r}\right) \frac{\partial}{\partial r}\ \ \ \mbox{and}\ \ \ \vec l^-=- e^{-\beta} \frac{\partial}{\partial r}
\end{equation}
The expansion corresponding to the outgoing null geodesics takes the expression
\[
\theta^+= \frac{e^{ \beta}}{r} \left( 1-\frac{2 m}{r}\right).
\]
confirming that, as is well-known, a round sphere (defined by constant values of $u$ and $r$) in which $r=2 m$ is a MTS so that the hypersurface $r=2 m$ defines the apparent 3-horizon A3H. Note that the metric (\ref{SEEF}) is not singular on A3H, so that the used coordinates are a good choice (although not at all unique) for describing the tunneling of particles through marginally trapped round spheres.

Since we are dealing with compact round spheres we can compute the effective surface gravity along the A3H as
\[
\kappa= -\sqrt{\frac{A(\mathcal S_{A3H})}{16\pi}}\  l^{-\alpha}\nabla_\alpha \theta^+\rfloor_{A3H}=\left.\frac{1-2 m' }{4 m}\right\rfloor_{A3H},
\]
where $m'$ stands for the partial derivative of $m$ with respect to $r$.
Therefore, the emission rate satisfies
\begin{equation}\label{ERD}
\Gamma \sim \exp \left(-8 \pi \left.\frac{ m}{1-2 m'}\right\rfloor_{A3H} \omega_{A3H}\right),
\end{equation}
where now $\omega_{A3H}$ is the value of $\omega$ on the A3H (which in this dynamical case is not a constant, but depends on the specific marginally trapped round sphere on the A3H we are considering).
It is also clear that the \textit{effective temperature} (when applicable) takes the form
\[
T=\left.\frac{1-2 m' }{8 \pi m}\right\rfloor_{A3H}.
\]
If one wants to know the \emph{temperature} of the MTS as measured by some observers that do detect tunneling radiation, we must first choose the specific observer and study the geodesics that transport the emitted radiation up to her.
To exemplify this and in order to compare with well-known results, let us obtain the temperature as measured by a static observer at infinity in the Schwarzschild black hole in which $m=M$ is a constant and $\beta=0$. As already pointed out by the QFT treatment, the existence of the Killing vector $\vec \xi =\partial_{u}$ defines a family of privileged (static) observers of the radiation outside the BH and, in particular, at infinity. On the other hand, this is the simplest case to treat since the associated constant of motion for geodesics $E=-S_{0,\alpha} \xi^\alpha=-S_{0,u}$ (corresponding to the energy of the radiation as measured by static observers at infinity) makes unnecessary any further work with the geodesics. At infinity, $\omega= E$ so that, from (\ref{ERD}),
\begin{equation}\label{ERS}
\Gamma \sim \exp \left(-8 \pi M E\right),
\end{equation}
which can be compared with the emission rate, in case of thermal emission, as measured from infinity  $\Gamma \sim \exp (-E/T)$ in order to deduce that a static observer at infinity would measure a temperature for the emitted radiation
\[
T=\frac{1}{8\pi M}.
\]
This corresponds with the standard result by Hawking and, for this case, also with our effective temperature.

\begin{figure}
\centering
\includegraphics[scale=.9]{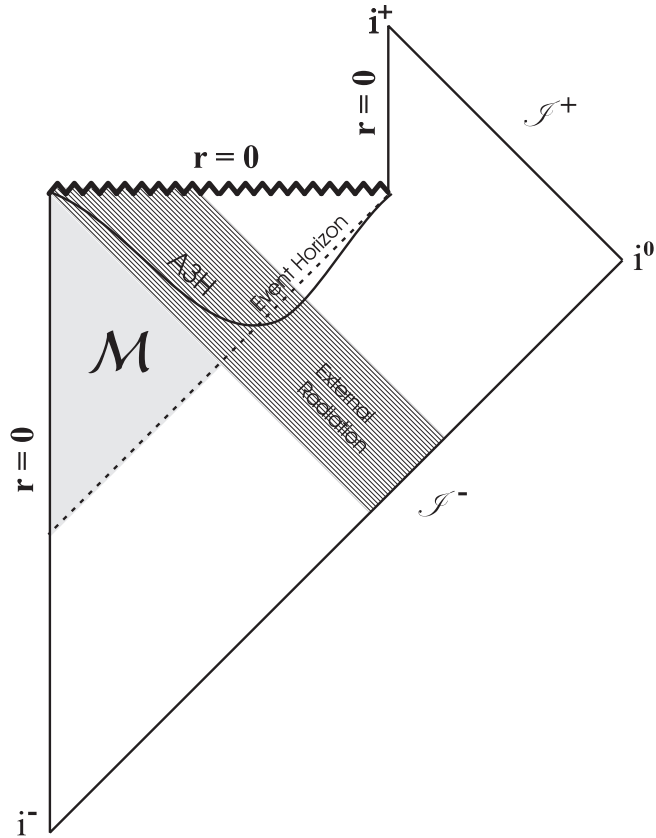}
\caption{\label{figX} An example of a (semiclassical) dynamical black hole which is generated by the implosion of external radiation and eventually evaporates. Particle production can be associated with the A3H ($r=2 M(u)$; solid line), but also with MTS in the spacetime according with some privileged observers. Thus, for example, the privileged observers expect radiation to be generated and to travel through some regions of the Minkowskian grey region $\cal M$. Of course, they also expect radiation traveling towards the future null infinity ---thus causing the evaporation of the black hole.}
\end{figure}

A simple dynamical model representing the generation of a black hole and its later evaporation can be described by considering the collapse of incoherent radiation and using the Vaidya solution in advanced coordinates --- what corresponds to the case $\{\beta=0,\ m=M(u) \}$. A particular case for a specific $M(u)$ has been drawn in figure \ref{figX}. We could now calculate the temperature for round spheres on the A3H (in the places where the evolution were slow enough) as measured by specific privileged observers in this spacetime, however we have drawn this figure with a different goal. Specifically, our aim is to remind the reader that
other (non-spherically symmetric) MTS exist in this spacetime. In particular, the results in \cite{B&S0}\cite{B&S} (and references therein) suggest that, depending on the specific evolution of the mass function,  there may exist compact MTSs reaching some regions of the Minkowskian grey region $\mathcal M$. Our results in subsection \ref{GITF} would imply that there should be an effective surface gravity, an effective temperature and radiation also associated with these MTSs reaching the Minkowskian region. Furthermore, a perturbation argument in \cite{B&S} demonstrates that there are MTSs intersecting both sides of the A3H in every zone where the A3H is not null. All these MTSs will produce a certain amount of radiation. The question of which part of this radiation will eventually be seen at future null infinity depends on the specific model, and on how close to the event horizon the MTSs are.
Note that in the previous literature on \textit{Hawking radiation} only round spheres had been considered.

\subsection{Kerr-Vaidya Black Hole}

Let us now analyze a dynamical non-spherically symmetric case of possible astrophysical relevance: The Kerr-Vaidya solution. Its line-element can be written in advanced Eddington-Finkelstein coordinates\footnote{The line-element in retarded Eddington-Finkelstein coordinates ($u\leftrightarrow -u$) was first published in \cite{MT}.} as:
\begin{eqnarray*}
ds^2&=&-\left(1-\frac{2 M(u) r}{\rho^2}\right) du^2+ 2\ du\ dr+ \rho^2 d\theta^2-\frac{4 a M(u) r \sin^2\theta}{\rho^2} d\phi\ du\\
& &-2 a \sin^2\theta\ d\phi\ dr + \frac{(r^2+a^2)^2-a^2 \Delta \sin^2\theta}{\rho^2}\sin^2\theta d\phi^2,
\end{eqnarray*}
where $\rho^2\equiv r^2+a^2 \cos^2\theta$ and
$\Delta=\Delta(u,r)\equiv r^2-2 r M(u)+a^2$.
This is a non-vacuum solution of Einstein's field equations which contains matter violating the weak energy condition ---unless $M(u)=M\neq 0$ is constant, in which case the metric reduces to that of the Kerr rotating black hole. For some discussion see \cite{MT,G}. It is more difficult to discuss the black hole nature of the space-time in the generic case with non-constant $M(u)$, because it is not easy to find any dynamical horizon (or more generally, any marginally trapped tube) even though the space-time may contain closed (marginally) trapped surfaces. (Readers are warned that some mistakes concerning this and other characteristics of this solution are spread in the literature).

We analyze the possibility of particle production associated with the surfaces $\mathcal S$ belonging to the family of compact surfaces (topologically $S^2$) given by constant values of the coordinates $u$ and $r$. (For a different approach relating particle emission to the \emph{event horizon}, see \cite{JW}). Two obvious normal one-forms are given by $du$ and $dr$, and then the null normals, normalized with $l^{+}_{\mu}l^{-\mu}=-1$, can be given by
$$
\mathbf{\underline{l}}^+ = \frac{\rho^2}{2 \Theta^2} \left(-\Delta du +(r^2+a^2+ \Theta) dr \right) , \hspace{1cm} \mathbf{\underline{l}}^- = -du +\frac{r^2+a^2- \Theta}{\Delta} dr
$$
where $\Theta$ is a shorthand for
$$
\Theta = \sqrt{(r^2+a^2)\rho^2+2Mra^2\sin^2\theta} .
$$
Observe that at the roots of $\Delta =0$ one also has $r^{2}+a^{2}-\Theta =0$ and thus the $dr$-component of $\mathbf{\underline{l}}^-$ can be well defined even if those roots exist (what happens whenever $M^2(u) \geq a^2$).

The mean curvature 1-form can be computed (e.g. using the formulas in \cite{S}) to get
$$
\mathbf{\underline{H}}=\frac{1}{\Theta^{2}} \left[[2r\rho^2+a^2\sin^2\theta (M+r)] dr +a^2r \dot M \sin^2\theta\,   du \right]
$$
where $\dot M=dM/du$ represents the derivative of $M(u)$. From here one can easily get the null expansions
\begin{eqnarray*}
\theta^+ =\frac{1}{2 \Theta^3}\left[ \Delta (2r\rho^2+a^2\sin^2\theta (M+r))+ \dot M r a^2\sin^2\theta (r^2+a^2+ \Theta)\right] , \\
\theta^-=-\frac{1}{\Theta \rho^2}\left[ 2r\rho^2+a^2\sin^2\theta (M+r)+ \frac{r^2+a^2- \Theta}{\Delta} \dot M r a^2\sin^2\theta \right]
\end{eqnarray*}
and the dual expansion vector
\begin{eqnarray}
\ast \vec H =\frac{1}{\rho^{2}\Theta}\left\{[2r\rho^2+a^2\sin^2\theta (M+r)] \frac{\partial}{\partial u}- \dot M r a^{2}\sin^{2}\theta \frac{\partial}{\partial r} \right. \nonumber \\
\left. -\frac{a}{\Theta^2} \left[\rho^{2}\dot M r a^{2}\sin^{2}\theta-2Mr[2r\rho^2+a^2\sin^2\theta (M+r)]  \right] \frac{\partial}{\partial \phi} \right\}.\label{DEK}
\end{eqnarray}

As we can see from the expression for $\theta^{\pm}$, the expansions cannot vanish on any entire surface in the family ---they can vanish somewhere on the surface, that is, for some values of $\theta$, but not elsewhere---, unless the surface lies on the intersection of $\Delta=0$ with $\dot M =0$. Thus, if there is a value $u=u_{0}$ such that
$$
\dot M(u_{0})=0 , \hspace{1cm} M^2(u_{0})Ê\geq a^{2}
$$
then the expansion $\theta^{+}$ vanishes on the entire surfaces given by $u=u_{0}, r= r_{\pm}$, where $r_{\pm}$ are the roots of $\Delta (u_{0},r)=0$:
$$ r_{\pm}= M(u_{0})\pm \sqrt{M^2(u_{0})-a^2}.
$$

In general, these surfaces do not foliate any dynamical horizon (or marginally trapped tube), except for the Kerr case with $M =$ constant , in which case the hypersurface $\Delta =0$ is the event horizon ---and a Killing horizon.

For concreteness, we will focus on radiation from an outer MTS $(u_0,\ r_{+})$ (although one can proceed in a similar manner for an inner MTS $(u_0, r_{-})$).
Since the surfaces defined by constant values of $u$ and $r$ are compact, we can write the effective surface gravity (\ref{ESG}) associated with these MTSs as
\begin{eqnarray*}
\kappa&=& -\sqrt{\frac{A(\mathcal S_{+})}{16\pi}}\  l^{-\alpha}\nabla_\alpha \theta^+\rfloor_{+}=\\
&=&\frac{(r_+ - M) [4 r_+^2 \rho_+^2+a^2(a^2+3 r_+^2)\sin^2\theta]+a^4 r_+^2 \sin^4\theta\ \ddot M_+}{4 r_+ (a^2+r_+^2)^{3/2} \rho_+^2},
\end{eqnarray*}
where the subscript `$+$' indicates that the quantity is evaluated on the MTS $(u_0, r_+)$ and we have used that the area of the MTS is $A(\mathcal S_{+})=4\pi (r_{+}^2+a^2)$.

The emission rate will then satisfy
\begin{equation}\label{emrakerr}
\Gamma \sim \exp \left(-8 \pi \frac{r_+ (a^2+r_+^2)^{3/2} \rho_{+}^2 \omega_{+}}{(r_+ - M) [4 r_+^2 \rho_+^2+a^2(a^2+3 r_+^2)\sin^2\theta]+a^4 r_+^2 \sin^4\theta\ \ddot M_+}\right).
\end{equation}

Note that both the effective surface gravity and the emission rate in the general Kerr-Vaidya solution depend on the value of the coordinate $\theta$.

Obtaining the \textit{effective temperature} is straightforward by using the results in subsection \ref{compactMTS}. However, to compute the \emph{temperature} ---associated with the radiation measured by one of the observers that detects it ---requires identifying the specific observer and giving the function $M(u)$ explicitly.
To exemplify the procedure, we treat the simplest case with $M(u)=M=$constant: the Kerr solution. Then $\vec\xi_u\equiv\partial/\partial u$ and $\vec\xi_\varphi\equiv\partial/\partial \varphi$ are Killing vectors providing two constants of motion along the null geodesics
\begin{equation}\label{EJ}
E=-S_{0,\alpha} \xi_u^\alpha=-S_{0,u}\ \ \ \mbox{and}\ \ \ J=S_{0,\alpha} \xi_\varphi^\alpha=S_{0,\varphi}
\end{equation}
that notably simplifies dealing with them. We also choose, from our privileged observers (with unitary 4-velocity $\hat\zeta$ parallel to $\ast \vec H $), the particular ones at infinity. They measure an energy for a particle radiated from the MTS
\[
\varepsilon=-S_{0,\alpha}\hat\zeta^\alpha\rfloor_\infty=E.
\]
To identify the temperature $T$ we take into account the statistical-mechanics fact that, according to these observers, the argument in the exponential of (\ref{emrakerr}) should take the form $\alpha-\varepsilon/T$.
By using the definition of $\omega$ (\ref{EEnergy}) together with (\ref{DEK}) and (\ref{EJ}) we get
\begin{equation}\label{OK}
\omega_{+}= \frac{2 r_+ \rho_+^2+a^2\sin^2\theta (M+r_+)}{2\rho_+^2\sqrt{a^2+r_+^2} } (E-\Omega_{+} J ),
\end{equation}
where
\[
\Omega_{+}=\frac{a}{r_{+}^2+a^2}
\]
is the (constant) \textit{angular velocity of the horizon} \cite{Wald}.
If we rewrite (\ref{emrakerr}) using (\ref{OK}) we get
\[
\Gamma \sim \exp \left[-\frac{\varepsilon-\Omega_{+} J}{T}\right],
\]
that allows us to get the temperature measured at infinity to be
\[
T=\frac{\sqrt{M^2-a^2}}{4\pi M (M+\sqrt{M^2-a^2})}.
\]
Note that this temperature does not depend of the angular variables. Both the emission rate and the temperature for this particular Kerr-metric case coincide with previous results in QFT \cite{Parker}.

\subsection{Cosmological LRS Bianchi I Horizon}\label{BI}
As a final example, consider now a cosmological situation, given by the following locally rotationally symmetric (LRS) Bianchi I spacetimes
$$
ds^2 = -dt^2 + a^2(t)(dx^2+dy^2) + b^2(t) dz^2
$$
where $a(t)$ and $b(t)$ are arbitrary functions of the cosmological time $t$. The particular case $a=b$ is obviously the spatially flat Robertson-walker model with scale factor $a(t)$. Consider now the following change of coordinates
$$
r=\sqrt{x^2+y^2+z^2}, \hspace{1cm} \tan \vartheta =\frac{\sqrt{x^2+y^2}}{z}, \hspace{1cm} \tan\varphi =\frac{y}{x}
$$
which brings the metric to the form
\begin{eqnarray*}
ds^2 = -dt^2+\left(a^2\sin^2\vartheta+b^2\cos^2\vartheta \right)dr^2+2r(a^2-b^2)\sin\vartheta\cos\vartheta dr d\vartheta \\+r^2\left(a^2\cos^2\vartheta+b^2\sin^2\vartheta \right)d\vartheta^2 +a^2 r^2 \sin^2\vartheta d\varphi^2
\end{eqnarray*}
and the family of compact 2-dimensional surfaces
defined by constant values of $t=t_0$ and $r=r_0$. A moderately long but straightforward calculation (using for instance the formulas in \cite{S} ) leads to the following expression for the mean curvature one-form
$$
H_\mu dx^\mu =
\left(\frac{\dot a}{a} +\frac{a\dot a \cos^2\vartheta +b \dot b \sin^2\vartheta}{a^2\cos^2\vartheta+b^2\sin^2\vartheta } \right)dt + \frac{b^2\left[a^2(1+\cos^2\vartheta)+b^2\sin^2\vartheta\right]}{r\left(a^2\cos^2\vartheta+b^2\sin^2\vartheta \right)^2}dr
$$
from where, together with the expressions for the two null vector fields orthogonal to the family $\{S_{t_0,r_0}\}$
$$
\sqrt{2}\, \vec l^\pm = \partial_t \pm \frac{\left(a^2\cos^2\vartheta+b^2\sin^2\vartheta \right)^{1/2}}{ab}\partial_r \mp \frac{(a^2-b^2)\sin\vartheta\cos\vartheta}{abr \left(a^2\cos^2\vartheta+b^2\sin^2\vartheta \right)^{1/2}}\partial_\vartheta
$$
one can readily get the two null expansions on each $S_{t_0,r_0}$
$$
\sqrt{2}\, \theta^\pm =\left. \left(\frac{\dot a}{a} +\frac{a\dot a \cos^2\vartheta +b \dot b \sin^2\vartheta}{a^2\cos^2\vartheta+b^2\sin^2\vartheta } \right)\pm \frac{b\left[a^2(1+\cos^2\vartheta)+b^2\sin^2\vartheta\right]}{ar_0\left(a^2\cos^2\vartheta+b^2\sin^2\vartheta \right)^{3/2}} \right|_{t=t_0} \, .
$$
Assume then that there is a slice $t=t_0$ such that
\begin{equation}
a(t_0) =b(t_0) , \hspace{1cm} \dot a(t_0) = \dot b(t_0)\neq 0 .\label{cond}
\end{equation}
Then, by choosing $r_0$ such that
$$
|\dot a(t_0)| =\frac{1}{r_0}
$$
the surface $t=t_0$ and $r=r_0$ is marginally trapped, future (past) trapped if the Universe is contracting (expanding) at $t_0$. This is not surprising, as the Universe is instantaneously RW at $t_0$ and the surface is then a typical MTS, observe however that the functions $a(t)$ and $b(t)$ are fully arbitrary, apart from (\ref{cond}). One can also check that in the immediate neighborhood of this MTS there are trapped and untrapped surfaces in the chosen family.

In order to analyze the instantaneous emission of particles from such MTS and for concreteness, let us consider the case of a Universe expanding at $t_0$ (the contracting case can be computed similarly), what implies that $\theta^+\rfloor_{MTS}>0$ and $\theta^-\rfloor_{MTS}=0$. Note that with this choice, contrarily to the previous examples, we will be dealing with a marginally \emph{past} trapped surface. Only for this example, we will not rename the above quantities (i.e., change the superscripts `+'$\leftrightarrow$`-') to emphasize this fact.
Another important difference in this case is that a negative-$\omega$ particle tunneling the MTS passes from an \textit{interior} region with $\theta^-<0$ to an \textit{exterior} region with $\theta^->0$. In this way, the sign changes in (\ref{pathint}) so that, for the case of marginally \emph{past} trapped surfaces is convenient to define the effective surface gravity as
\[
\kappa\equiv +\sqrt{\frac{A(\mathcal S_{MTS})}{16\pi}}\  l^{+\alpha}\nabla_\alpha \theta^-\rfloor_{MTS}.
\]
For our particular case we have
\begin{equation}\label{kappapast}
\kappa=\frac{2+r_0^2 a(t_0) [(1+\cos^2\vartheta)\ \ddot a(t_0)+\sin^2\vartheta\ \ddot b(t_0)]}{4 r_0 a(t_0)}
\end{equation}
Therefore, the corresponding emission rate will be
\begin{equation}
\Gamma \sim \exp \left(\frac{-8 \pi\  r_0 a(t_0)\ \omega\rfloor_{MTS}}{2+r_0^2 a(t_0) [(1+\cos^2\vartheta)\ \ddot a(t_0)+\sin^2\vartheta\ \ddot b(t_0)]}\right).
\end{equation}
Note that, as previously mentioned, if the cosmological solution satisfied $a(t)=b(t)$, we would be dealing with the particular case of a flat Robertson-Walker solution in which case
\[
\Gamma \sim \exp \left(\frac{-4 \pi\ \dot a(t) a(t)\ \omega\rfloor_{H}}{\dot a^2(t)+ a(t) \ddot a(t)}\right)
\]
on the marginally trapped tube defined by $r=|\dot a|^{-1}$. This expression coincides with previously found results for the general RW solutions ---which had been already treated in the literature due to its spherical symmetry (around every point), what had allowed the use of the \textit{Kodama vector approach} (see \cite{revVz}).

Following the procedure explained in the previous examples we can now, for example, find the temperature that a privileged observer co-moving with the cosmological fluid at $r=0$ would measure for the radiation. Taking into account that this observer measures an energy for the particles $E=-S_{0,t}=\omega$, the straightforward result is that, in the case of the de Sitter space, this temperature takes the form
\[
T=\frac{\mathcal H}{2\pi},
\]
where $\mathcal H\equiv\dot a/a$. This is a well-known result obtained in the literature
by using many different approaches. However, in a general RW spacetime $\dot{\mathcal H}$ plays an active role in the measured temperature since we get
\[
T=\frac{\kappa}{2\pi}=\frac{2 \mathcal{ H}^2+\dot {\mathcal H}}{4 \pi \mathcal H}.
\]

\section{Conclusions}\label{Conclu}

In this paper we have presented an invariant formalism for the computation of particle production through tunneling as perceived by some privileged observers. We have shown that the perception of this radiation can be associated with general marginally trapped surfaces belonging to general spacetimes. In the process, a definition of effective surface gravity for compact MTS has naturally arisen. Our approach requires the definition of a family of surfaces $\{\mathcal S(t)\}$ with respect to which the computations are carried out. In other words, there is a family of privileged observers (associated with the untrapped surfaces of the family) with 4-velocity directed in the direction of the dual expansion vector $\ast \vec H$  with respect to whom one defines the effective magnitudes related to the phenomenon. Each of these observers sees her corresponding $\{\mathcal S(t)\}$ as {\em non-expanding}.
On the other hand, for a vector field $\ast \vec H$ defined in a \emph{region} of the spacetime and a canonical time chosen in accordance with the privileged observers associated with that vector field,  there will be a preferred vacuum state and a privileged definition of {\it particles} in the region. In this way one can safely analyze the particles arisen via tunneling as perceived or detected by these privileged observers \emph{at any point in the region}. Of course, as it is well-known from QFT, other unrelated observers will choose a different preferred vacuum state and a different definition of particle, so that they can have a completely different description of the particle production process, what includes that they could even declare its inexistence. For example, in the case of the dynamical spherically symmetric solutions (sec.\ref{SE}) we have only treated specific privileged observers associated with regions where the round spheres are untrapped so that, in fact, we have been dealing with a specific canonical time, vacuum state and particle definition as described only by those observers \footnote{Note that in the paper we have made explicit only the particular observations of an observer at infinity in Schwarzschild's spacetime, but the observations of any other of our privileged observers could be also described in any general spherically symmetric spacetime.}.

We have checked our formalism with examples of astrophysical relevance by analyzing the tunneling through particular MTS. This includes the case of the above mentioned dynamical spherically symmetric solutions as well as the case of the non-spherically symmetric Kerr-Vaidya solution. We have also checked the formalism with a cosmological example represented by a locally rotationally symmetric Bianchi I solution. In this way, we have been able to see, in practice, that the formalism reproduces previously known results and that it allows us to obtain the physical magnitudes associated with the radiation for new cases and, ultimately, in the general case.

Some authors had previously considered stationary and/or dynamical spherically symmetric spacetimes and had associated \textit{Hawking radiation} with marginally trapped round spheres on A3H (and thus having $r=2m$).
However, in spherically symmetric situations other non-spherically symmetric marginally trapped surfaces do exist with points satisfying $r>2 m$ \cite{B&S}. Moreover, in some cases an observer could cross a marginally trapped surface at a point in which the mass function vanishes, even with her {\em entire} causal past being a piece of flat Minkowski spacetime \cite{B&S0}. Consider, for example, the case of imploding radiation in figure \ref{figX}. According to our results an observer crossing the region $\mathcal M$ could detect radiation even before the black hole has formed. This would be surprising if one identifies the tunneling horizon with $r=2m$, but it is not if one considers the existence of general non-spherical MTS in the region and that they are endowed with ``clairvoyance'' \cite{B&S0}\cite{B&S}.

Note finally that we have thoroughly neglected the back-reaction of the spacetime to the emission of radiation. In fact, we were forced to proceed in this way since only in transitions from spherically symmetric configurations to spherically symmetric configurations back-reaction can be easily and reliably computed due to the absence of gravitational radiation. Nevertheless, this does not seem a cause for concern since neglecting back-reaction is an usual approximation which one expects to be correct in astrophysical situations where closed trapped surfaces exist and whenever their enclosed Hawking energy is of astrophysical proportions so that they are far from their `total evaporation'. However, one should be aware that there are probably some interesting features that we are missing by being unable to consider back-reaction. For example, there could be small deviations from the thermal spectrum if the back-reaction could be taken into account. This has been shown for the simpler case of the tunneling through the spherically symmetric horizon in the Schwarzschild black hole solution \cite{P&W}. Moreover, back-reaction could take an important role in understanding the nature of the tunneling mechanism and its relationship with the standard phenomenon of tunneling through a barrier in quantum mechanics \cite{Parikh}.

\section*{Acknowledgements}
JMMS is supported by grants
FIS2010-15492 (MICINN), GIU12/15 (Gobierno Vasco), P09-FQM-4496 (J. Andaluc\'{\i}a---FEDER) and UFI 11/55 (UPV/EHU). R Torres acknowledges the financial support of the Ministerio de Econom\'{\i}a y Competitividad (Spain), projects MTM2014-54855-P. R Torres wishes to express his gratitude to JMMS for his hospitality during a stay in the Dept. of Theoretical Physics at the UPV/EHU.

\end{document}